\PassOptionsToPackage{unicode}{hyperref}
\PassOptionsToPackage{hyphens}{url}
\PassOptionsToPackage{dvipsnames,svgnames,x11names}{xcolor}

\documentclass[12pt]{article}
\usepackage{setspace}
\usepackage{mathrsfs}
\usepackage{natbib,graphicx,lscape,longtable}
\usepackage{mathrsfs,amsmath,amsthm,amssymb,color}
\usepackage{natbib,epsfig,graphicx,pdfpages}
\usepackage{rotating}
\usepackage{float}
\usepackage{bm}
\usepackage{caption}
\usepackage{subcaption}
\usepackage{ragged2e}
\usepackage[normalem]{ulem}
\usepackage{CJKutf8}
\usepackage{comment,booktabs,xcolor}
\usepackage[colorlinks=true, citecolor=black]{hyperref}

\usepackage{ragged2e}
\usepackage{algorithm}
\usepackage{algorithmic}
\usepackage{fontawesome5}

\usepackage{cleveref}
\bibpunct{(}{)}{;}{a}{,}{,}

\usepackage{iftex}
\ifPDFTeX
  \usepackage[T1]{fontenc}
  \usepackage[utf8]{inputenc}
  \usepackage{textcomp} % provide euro and other symbols
\else % if luatex or xetex
  \usepackage{unicode-math}
  \defaultfontfeatures{Scale=MatchLowercase}
  \defaultfontfeatures[\rmfamily]{Ligatures=TeX,Scale=1}
\fi
\usepackage{lmodern}
\ifPDFTeX\else
\fi
\IfFileExists{upquote.sty}{\usepackage{upquote}}{}
\IfFileExists{microtype.sty}{% use microtype if available
  \usepackage[]{microtype}
  \UseMicrotypeSet[protrusion]{basicmath} % disable protrusion for tt fonts
}{}
\makeatletter
\@ifundefined{KOMAClassName}{% if non-KOMA class
  \IfFileExists{parskip.sty}{%
    \usepackage{parskip}
  }{% else
    \setlength{\parindent}{0pt}
    \setlength{\parskip}{6pt plus 2pt minus 1pt}}
}{% if KOMA class
  \KOMAoptions{parskip=half}}
\makeatother
\usepackage{xcolor}
\makeatletter
\ifx\paragraph\undefined\else
  \let\oldparagraph\paragraph
  \renewcommand{\paragraph}{
    \@ifstar
      \xxxParagraphStar
      \xxxParagraphNoStar
  }
  \newcommand{\xxxParagraphStar}[1]{\oldparagraph*{#1}\mbox{}}
  \newcommand{\xxxParagraphNoStar}[1]{\oldparagraph{#1}\mbox{}}
\fi
\ifx\subparagraph\undefined\else
  \let\oldsubparagraph\subparagraph
  \renewcommand{\subparagraph}{
    \@ifstar
      \xxxSubParagraphStar
      \xxxSubParagraphNoStar
  }
  \newcommand{\xxxSubParagraphStar}[1]{\oldsubparagraph*{#1}\mbox{}}
  \newcommand{\xxxSubParagraphNoStar}[1]{\oldsubparagraph{#1}\mbox{}}
\fi
\makeatother

\usepackage{longtable,booktabs,array}
\usepackage{calc} % for calculating minipage widths
\usepackage{etoolbox}
\makeatletter
\patchcmd\longtable{\par}{\if@noskipsec\mbox{}\fi\par}{}{}
\makeatother
\IfFileExists{footnotehyper.sty}{\usepackage{footnotehyper}}{\usepackage{footnote}}
\makesavenoteenv{longtable}
\usepackage{graphicx}
\makeatletter
\def\maxwidth{\ifdim\Gin@nat@width>\linewidth\linewidth\else\Gin@nat@width\fi}
\def\maxheight{\ifdim\Gin@nat@height>\textheight\textheight\else\Gin@nat@height\fi}
\makeatother
\setkeys{Gin}{width=\maxwidth,height=\maxheight,keepaspectratio}
\makeatletter
\def\fps@figure{htbp}
\makeatother

\makeatletter
\@ifpackageloaded{caption}{}{\usepackage{caption}}
\AtBeginDocument{%
\ifdefined\contentsname
  \renewcommand*\contentsname{Table of contents}
\else
  \newcommand\contentsname{Table of contents}
\fi
\ifdefined\listfigurename
  \renewcommand*\listfigurename{List of Figures}
\else
  \newcommand\listfigurename{List of Figures}
\fi
\ifdefined\listtablename
  \renewcommand*\listtablename{List of Tables}
\else
  \newcommand\listtablename{List of Tables}
\fi
\ifdefined\figurename
  \renewcommand*\figurename{Figure}
\else
  \newcommand\figurename{Figure}
\fi
\ifdefined\tablename
  \renewcommand*\tablename{Table}
\else
  \newcommand\tablename{Table}
\fi
}
\@ifpackageloaded{float}{}{\usepackage{float}}
\floatstyle{ruled}
\@ifundefined{c@chapter}{\newfloat{codelisting}{h}{lop}}{\newfloat{codelisting}{h}{lop}[chapter]}
\floatname{codelisting}{Listing}

\makeatother
\makeatletter
\@ifpackageloaded{caption}{}{\usepackage{caption}}
\@ifpackageloaded{subcaption}{}{\usepackage{subcaption}}
\makeatother

\ifLuaTeX
  \usepackage{selnolig}  % disable illegal ligatures
\fi
\usepackage[]{natbib}
\usepackage{bookmark}

\IfFileExists{xurl.sty}{\usepackage{xurl}}{} % add URL line breaks if available
\hypersetup{
  pdftitle={Title},
  pdfauthor={Author 1; Author 2},
  pdfkeywords={3 to 6 keywords, that do not appear in the title},
  colorlinks=true,
  linkcolor={blue},
  filecolor={Maroon},
  citecolor={Blue},
  urlcolor={Blue},
  pdfcreator={LaTeX via pandoc}}

\newcommand{\anon}{0}

\newtheorem{theorem}{Theorem}
\newtheorem{lemma}{Lemma}
\newtheorem{proposition}{Proposition}
\newtheorem{corollary}{Corollary}

\def\beq{\begin{equation}}
\def\eeq{\end{equation}}
\def\beqr{\begin{eqnarray}}
\def\eeqr{\end{eqnarray}}
\def\beqrs{\begin{eqnarray*}}
\def\eeqrs{\end{eqnarray*}}
\def\bet{\begin{theorem}}
\def\eet{\end{theorem}}
\def\bel{\begin{lemma}}
\def\eel{\end{lemma}}
\def\bep{\begin{proposition}}
\def\eep{\end{proposition}}
\def\bg{\begin{figure}[tbph]\begin{center}}
\def\eg{\end{center}\end{figure}}

\def\bc{\begin{center}}
\def\ec{\end{center}}

\newtheorem{remark}{Remark}

\def\wt{\widetilde}

\def\mE{\mathcal E}

\def\mV{\mathcal{V}}

\def\bZ{\mathcal{Z}}

\def\mX{\mathbb{X}}

\def\mY{\mathbb{Y}}
\def\mZ{\mathbb{Z}}

\def\mI{\mathcal I}

\def\mE{\mathcal E}

\def\mV{\mathcal V}

\newcommand{\ve}{{\varepsilon}}

\def\bZ{\mbox{\boldmath $Z$}}

\usepackage{array}
\newcolumntype{H}{>{\setbox0=\hbox\bgroup}c<{\egroup}@{}}

\renewcommand{\arraystretch}{1.3}

\numberwithin{equation}{section}

\begin{document}

\def\spacingset#1{\renewcommand{\baselinestretch}%
{#1}\small\normalsize} \spacingset{1}

%%%%%%%%%%%%%%%%%%%%%%%%%%%%%%%%%%%%%%%%%%%%%%%%%%%%%%%%%%%%%%%%%%%%%%%%%%%%%%

\if0\anon
{
  \title{\bf A Propagation Framework for Network Regression}
  \author{Yingying Ma%\thanks{The authors gratefully acknowledge \textit{please remember to list all relevant funding sources in the version that gives all author information}}
  \hspace{.2cm}\\
    School of Economics and Management, Beihang University\\
   % and \\
    Chenlei Leng \\
     Department of Applied Mathematics, Hong Kong Polytechnic University}
     \date{}
  \maketitle
} \fi

\if1\anon
{
  \bigskip
  \bigskip
  \bigskip
  \begin{center}
    {\LARGE\bf A Propagation Framework for Network Regression}
\end{center}
  \medskip
} \fi

\bigskip
\begin{abstract}
We introduce a unified and computationally efficient framework for regression on network data, addressing limitations of existing models that require specialized estimation procedures or impose restrictive decay assumptions. Our Network Propagation Regression (NPR) models outcomes as functions of covariates propagated through network connections, capturing both direct and indirect effects. NPR is estimable via ordinary least squares for continuous outcomes and standard routines for binary, categorical, and time-to-event data, all within a single interpretable framework. We establish consistency and asymptotic normality under weak conditions and develop valid hypothesis tests for the order of network influence. Simulation studies demonstrate that NPR consistently outperforms established approaches, such as the linear-in-means model and regression with network cohesion, especially under model misspecification. An application to social media sentiment analysis highlights the practical utility and robustness of NPR in real-world settings.
\end{abstract}

\noindent%
\vfill

\newpage
\spacingset{1.65} % DON'T change the spacing!

%========================================
\section{Introduction}\label{sec:introduction}
Regression analysis forms the foundation of statistical modeling, typically assuming independent observations $(Y_i,x_i)$ drawn from
\[
\mY = \mX\bm{\beta} + \mE,
\]
where $\mY = (Y_1,\dots,Y_N)^\top$ denotes the response vector, $\mX=(x_1,\cdots,x_N)^\top$ is the design matrix, $\bm{\beta}$ is the vector of unknown regression coefficients, and $\mE$ represents  the error term. This independence assumption, however, fails for network-linked data, where outcomes for connected units influence one another through peer effects, spillovers, or spatial dependence.

We present a simple and flexible framework for network propagation regression (NPR), which models outcomes as functions of covariates diffused through a network over multiple steps. For continuous outcomes, NPR is a linear regression on propagated covariates:
\begin{equation}
\mY = \sum_{k=0}^{K} W^k \mX \bm{\lambda}_k + \mE, \label{network propagation model}
\end{equation}
where $W$ is the normalized network adjacency matrix, $W^k\mX$ represents covariates diffused $k$ steps through the network, and $\bm{\lambda}_k$ denotes the effect of covariates at network distance $k$. When $K = 0$, NPR reduces to standard regression, $\mY = \mX\bm{\lambda}_0 + \mE$. For $K \geq 1$, the additional terms $W^k\mX\bm{\lambda}_k$ capture network effects from paths of length $k$.

To estimate the parameters $\{\bm{\lambda}_k\}_{k=0}^K$, one simply applies ordinary least squares (OLS) via
\[
\texttt{lm}(\mY \sim \mX + W\mX + \cdots + W^K\mX),
\]
or equivalently, constructs the design matrix $\widetilde{\mX} = (\mX, W\mX, \ldots, W^K\mX)$ and computes $(\widetilde{\mX}^\top \widetilde{\mX})^{-1}\widetilde{\mX}^\top\mY$.

We establish consistency and asymptotic normality of the least-squares estimator under mild conditions, even when the true network dependence extends beyond any finite $K$ (i.e., as $K \to \infty$). Furthermore, we develop a hypothesis testing procedure to determine the
relevant propagation order from data.
The propagation structure generalizes naturally beyond linear regression. We develop two key extensions: \textbf{network logistic regression} for binary outcomes, which applies a logistic link to \eqref{network propagation model}, and \textbf{network Cox regression} for time-to-event outcomes, which incorporates propagated covariates into the hazard function. The same logic extends to all generalized linear models (GLMs), including count, multinomial, and censored outcomes, where only the link function changes, while the propagation architecture and estimation simplicity remain unchanged.

\subsection{Literature Review}

Network-linked data present a fundamental challenge to classical regression: observations are interdependent, with influence propagating through connections in the adjacency structure. An individual's outcome may be affected not only by its own characteristics but also by the traits of its neighbors--both direct and indirect. Ignoring these dependencies violates the independence assumption and can lead to biased estimation and invalid inference.

Econometricians have addressed this challenge through the \emph{linear-in-means (LIM)} model \citep{manski1993identification, bramoulle2009identification}, which specifies
\begin{equation}
\mY = \alpha\bm{1}_N + \rho W\mY + \mX\bm{\beta} + W\mX\bm{\delta} + \mE.\label{linear in mean model}
\end{equation}
In reduced form, LIM implies an infinite series $$\mY = \alpha(\mI_N-\rho W)^{-1}\bm{1}_N + \sum_{k=0}^\infty \rho^kW^k\mX(\rho\bm{\beta}+\bm{\delta}) + (\mI_N-\rho W)^{-1}\mE,$$  which does incorporate higher-order network effects, but in a highly restrictive geometric decay pattern ($\bm{\lambda}_k \propto \rho^k$). While mathematically elegant, this parametric form may be overly restrictive for applications where influence decays non-geometrically or varies by network distance.

The LIM framework also faces three additional challenges: (1) it is designed for continuous outcomes and does not naturally extend to binary, count, or survival responses; (2) the presence of $W\mY$ on the right-hand side introduces
simultaneity, complicating identification and requiring instrumental variables or generalized methods of moments (GMM); and (3) prediction requires inverting $(\mI_N - \rho W)$, which can be numerically unstable in large or ill-conditioned networks.

In parallel, statisticians have proposed \emph{regression with network cohesion (RNC)} \citep{li2019prediction}:
\begin{equation}
\mY = \bm{\mu} + \mX \bm{\beta} + \mE,
\label{RNC}
\end{equation}
where $\bm{\mu} = (\mu_1, \ldots, \mu_N)^\top$ is a vector of node-specific effects regularized by a graph Laplacian penalty. While RNC improves prediction through smoothing, it does not explicitly parameterize network effects, offers limited interpretability for diffusion processes, and requires estimating $N$ additional parameters, raising computational barriers for large networks.

Meanwhile, machine learning has developed powerful \emph{graph neural networks (GNNs)} such as Graph Convolutional Networks \citep{kipf2017semi} and their simplified variants \citep{wu2019simplifying}. These models iteratively aggregate neighborhood information to learn node representations, achieving strong predictive performance. However, they operate as black boxes: their parameters lack statistical interpretation, they do not provide inferential tools for peer effects, and they are prone to issues like over-smoothing without clear theoretical safeguards \citep{oonograph2020}.

The literature thus presents a dichotomy: structural models like LIM offer interpretability and identification but are rigid and limited to linear settings; RNC improves flexibility but obscures network mechanisms; GNNs deliver predictive power but sacrifice statistical rigor. What remains missing is a unified, interpretable, and estimable framework that explicitly models network diffusion, accommodates diverse response types, and scales to large networks while retaining theoretical guarantees.

Our network propagation regression (NPR) framework bridges these gaps. By modeling outcomes as a function of covariates propagated through the network, NPR retains the interpretability of structural models, generalizes naturally beyond continuous responses, avoids the simultaneity of LIM, and remains as simple to estimate as ordinary least squares. In the next section, we formalize the NPR approach and establish its theoretical properties.

\subsection{Contributions}

This paper makes the following contributions:

\noindent
\textbf{A unified propagation framework}
We introduce a propagation framework for network regression that models outcomes as a function of covariates diffused through the network. The approach generalizes naturally from continuous responses (via OLS) to binary, categorical, and time-to-event outcomes (via GLM links), while maintaining a single interpretable architecture.

\noindent
\textbf{Simplicity and estimability}
Unlike simultaneous-equation models that require instrumental variables or GMM, our framework propagates only exogenous covariates, avoiding endogeneity concerns. Estimation reduces to ordinary least squares (for continuous outcomes) or standard GLM routines, making implementation straightforward even for large networks.

\noindent
\textbf{Theoretical guarantees}
We establish consistency and asymptotic normality of the estimators under weak conditions that allow for dependent errors and do not require specific distributional assumptions. We also develop valid hypothesis tests for determining the effective radius of network influence.

\noindent
\textbf{Flexibility in propagation structure}
Our approach relaxes the restrictive geometric decay imposed by linear-in-means models, allowing network effects $\bm{\lambda}_k$ to follow any pattern across distances $k=0,1,\dots,K$. When propagation decays sufficiently fast, a finite $K$ provides a principled approximation with guaranteed error bounds.

\noindent
\textbf{Bridging interpretability and performance}
The framework generates structure-enhanced features $\{W^k\mX\}$ that can be fed into any machine learning algorithm, offering a model-agnostic way to inject network information while maintaining the interpretability of propagation coefficients $\bm{\lambda}_k$.

\noindent
\textbf{Empirical validation}
Numerical results  show our method outperforms existing approaches--including linear-in-means, regression with network cohesion, and graph neural networks--across diverse network structures and response types. An application to social media sentiment analysis demonstrates practical utility and robustness.

The paper is organized as follows. Section~\ref{sec:method} introduces the propagation framework and establishes its theoretical properties. Section~\ref{sec:GLM} extends it to binary and survival outcomes. Section~\ref{sec:simulation} presents simulation evidence, and Section~\ref{sec:real_data} illustrates the method on social media data. Section~\ref{sec:conclusion} concludes. Proofs and additional details appear in the Appendix.

\noindent
\textbf{Notations.}
Throughout, $\bm{0}_q$ and $\bm{0}_{p \times q}$ denote zero vectors and matrices, and $\mI_q$ is the $q \times q$ identity matrix. For $\bm u \in \mathbb{R}^p$, $\|\bm u\|_1$, $\|\bm u\|_2$, and $\|\bm u\|_4$ are the $\ell_1$, $\ell_2$, and $\ell_4$ norms. For a matrix $A$, $\lambda_{\max}(A)$ and $\lambda_{\min}(A)$ are its largest and smallest eigenvalues, $|A|$ its determinant, $\|A\|_F$ its Frobenius norm, and $\|A\|_2$ its operator norm; $a_{ij}$ is the $(i,j)$-entry. All vectors are column vectors unless noted. For sequences $a_N$ and $b_N$, $a_N = O(b_N)$ means $|a_N| \leq C|b_N|$ for some $C > 0$, and $a_N = o(b_N)$ means $a_N/b_N \to 0$ as $N \to \infty$. Convergence in probability and distribution are denoted $\xrightarrow{p}$ and $\xrightarrow{d}$, respectively.

 %========================================

%===============================================
\section{The Network Propagation Model} \label{sec:method}
\subsection{Model Formulation}\label{model1}
We formalize the Network Propagation Regression (NPR) for a directed network with $N$ nodes. Let $A \in \mathbb{R}^{N\times N}$ be the adjacency matrix, where $a_{ij} = 1$ if there is a directed edge from node $i$ to node $j$ ($a_{ii}=0$), and $a_{ij} = 0$ otherwise. For each node, we observe a response $Y_i$ and a $d$-dimensional covariate vector $x_i$. Let $\mY = (Y_1, \ldots, Y_N)^\top$ and $\mX = (x_1, \ldots, x_N)^\top$ denote the response vector and design matrix, respectively.

The central principle of NPR is that a node's outcome reflects not only its own covariates, but also the cumulative influence of covariates propagated through the network. To capture this, let $W$ be the row-normalized adjacency matrix derived from $A$, and define $W^k$ as the $k$-th power of $W$, where $w_{ij}^{(k)}$ quantifies the strength of $k$-step paths from node $i$ to node $j$. The model is specified as
\begin{equation}
\mY = \eta \bm{1}_N + \sum_{k=0}^{\infty} W^{k} \mX \bm{\lambda}_k + \mE, \label{infty_model}
\end{equation}
where $\eta$ is an intercept, $\bm{1}_N$ is the $N$-vector of ones, $\bm{\lambda}_k = (\lambda_{k1}, \ldots, \lambda_{kd})^\top$ are propagation coefficients for the $k$-step covariates, and $\mE = (\varepsilon_1, \ldots, \varepsilon_N)^\top$ is a mean-zero error vector. Multiplying both sides of \eqref{infty_model} by $\mI_N - N^{-1} \bm{1}_N \bm{1}_N^\top$ removes the intercept $\eta$, so we assume $\eta = 0$ henceforth.

We allow $\mE$ to have a general covariance structure, $\mathrm{Cov}(\mE) = UU^\top$, where $U$ is an arbitrary $N \times N$ matrix. This accommodates both independent and dependent errors, reflecting the possibility that unobserved factors may be correlated across nodes due to latent network effects or shared environments. Notably, our theoretical results do not require $\mE$ to be independent or identically distributed, making the framework robust to a wide range of dependence structures commonly encountered in network data.

The infinite sum in~\eqref{infty_model} captures the full diffusion process, but its convergence requires that propagation effects decay sufficiently rapidly, as formalized in Condition (C4) below. This guarantees both model stability and interpretability, as the influence of distant nodes becomes negligible. For practical implementation, we approximate the infinite series by a finite expansion as in \eqref{network propagation model}, retaining terms up to order $K$. This finite-order representation strikes a balance between model complexity and fidelity to the underlying diffusion process, and is a well-established approach in statistics and signal processing for approximating infinite series. The choice of $K$ determines the effective radius of network influence and can be selected empirically, as discussed in subsequent sections.

This formulation offers several advantages. First, it captures higher-order network effects: the term $W^{k} \mX$ incorporates covariates from nodes at path distance $k$, with $\bm{\lambda}_k$ quantifying their influence. Second, by propagating only exogenous covariates $\mX$--rather than endogenous responses $\mY$--the model avoids the simultaneity and identification issues common in traditional social interaction models. Third, estimation remains tractable: parameters can be efficiently estimated via ordinary least squares, with forward selection used to address multicollinearity among $(\mX, W\mX, \ldots, W^K \mX)$.

A key strength of NPR is its flexibility in modeling complex, higher-order dependencies, allowing a node's response to be influenced by covariates from all neighbors within its $K$-step neighborhood. This generalizes classical social interaction models, which typically consider only first-order effects.

A central theoretical question is under what conditions the truncated model~\eqref{network propagation model} provides asymptotically valid inference for the true process~\eqref{infty_model}. We show that the ordinary least squares estimators for $\bm{\lambda}_k$ ($k = 0, \ldots, K$) are consistent for their population counterparts, provided $K$ grows appropriately with sample size. The next section formalizes these results and establishes the asymptotic properties of the estimator under mild regularity conditions.

\subsection{Estimation}
\label{sec:estimation}

We first describe the estimation procedure for the case where the truncation parameter $K$ is fixed and known, deferring the data-driven selection of $K$ to Section~\ref{sec:test}. Let the augmented design matrix be $\widetilde{\mX} = (\mX, W\mX, \ldots, W^K\mX) \in \mathbb{R}^{N \times \widetilde{d}}$, where $\widetilde{d} = d(K+1)$, and define $\wt{\bm x}_i = \widetilde{\mX}^\top \bm{e}_i \in \mathbb{R}^{\widetilde{d}}$, with $\bm{e}_i$ denoting the $i$th canonical basis vector in $\mathbb{R}^N$.
As $K$ increases, the columns of $\widetilde{\mX}$ may become highly collinear due to repeated propagation through the network matrix $W$. To ensure numerical stability and identifiability, we remove linearly dependent columns, guaranteeing that $\widetilde{\mX}^\top \widetilde{\mX}$ is well-conditioned and invertible.

Assuming $N \gg \widetilde{d}$, the ordinary least squares (OLS) estimator for the parameter vector $\bm{\theta} = (\bm{\lambda}_0^\top, \bm{\lambda}_1^\top, \ldots, \bm{\lambda}_K^\top)^\top \in \mathbb{R}^{\widetilde{d}}$ is given by
\begin{equation}
\widehat{\bm{\theta}} = (\widetilde{\mX}^\top \widetilde{\mX})^{-1} (\widetilde{\mX}^\top \mY).
\end{equation}
This estimator is computationally efficient and leverages the linear structure of the model, facilitating both inference and interpretation. %The treatment of multicollinearity and the selection of relevant propagation terms are further discussed in Section~\ref{sec:selection}.

To establish the theoretical properties of $\widehat{\bm{\theta}}$, we impose the following regularity conditions. Let $a_N^2 = \max_{k \geq 1} \lambda_{\max}\big((W^{k})^\top W^k\big) \cdot \frac{\lambda_{\max}(N^{-1} \widetilde{\mX}^\top \widetilde{\mX})}{\lambda_{\min}(N^{-1} \widetilde{\mX}^\top \widetilde{\mX})}$, and let $\widetilde{\mX}_j$ denote the $j$-th column of $\widetilde{\mX}$.

\begin{itemize}
    \item[(C1)] {Error Structure.} Let $\{v_i\}_{i=1}^N$ be independent and identically distributed random variables with $\mathbb{E}[v_i] = 0$ and $\mathrm{Var}(v_i) = 1$. Assume $\mE = U \mV$, where $\mV = (v_1, \ldots, v_N)^\top$, and there exist constants $\delta_1 > 0$, $c_{\min,\mE} > 0$, and $c_{\max,\mE} < \infty$ such that $\mathbb{E}[|v_i|^{2+\delta_1}] = O(1)$ and $c_{\min,\mE} \leq \lambda_{\min}(U U^\top) \leq \lambda_{\max}(U U^\top) \leq c_{\max,\mE}$.

    \item[(C2)] {Dimensionality.} As $N \to \infty$, the total number of predictors satisfies $\widetilde{d} = O(N^{\alpha_1})$ for some $\alpha_1 \in (0,1)$.

    \item[(C3)] {Design Matrix.} (i) There exist constants $c_{\min} > 0$ and $c_{\max} < \infty$ such that $c_{\min} \leq \lambda_{\min}(N^{-1} \widetilde{\mX}^\top \widetilde{\mX}) \leq \lambda_{\max}(N^{-1} \widetilde{\mX}^\top \widetilde{\mX}) \leq c_{\max}$. \newline
    (ii) The fourth moment of the projected design satisfies $\max_{1 \leq j \leq \widetilde{d}} \|U^\top \widetilde{\mX}_j\|_4^4 = O(N^{\alpha_2})$, with $2\alpha_1 + \alpha_2 < \min\left\{\frac{2 + 3\delta_1}{2 + \delta_1}, 2\right\}$.

    \item[(C4)] {Truncation Error.} As $N \to \infty$ and $K \to \infty$, the omitted propagation terms satisfy $\sum_{k=K+1}^{\infty} \|\bm{\lambda}_k\| = o(a_N^{-1})$.
\end{itemize}

Our regularity conditions play distinct and complementary roles in establishing the theoretical properties of the OLS estimator. Condition (C1) substantially relaxes conventional i.i.d.\ Gaussian or sub-Gaussian error assumptions by allowing $\mathrm{Cov}(\mE) = UU^\top$, which accommodates arbitrary error correlation structures when $UU^\top$ is non-diagonal. This flexible framework, also adopted in \citet{Zhang2024}, encompasses a wide range of empirically relevant error processes, provided suitable moment conditions are maintained.

Condition (C2) permits the parameter dimension $\widetilde{d}$ to diverge with $N$ at a sublinear rate, $O(N^{\alpha_1})$ for $\alpha_1 < 1$.

Condition (C3) ensures the design matrix $\widetilde{\mX}$ is well-conditioned. Part (i) guarantees identifiability and numerical stability by bounding the eigenvalues of $N^{-1}\widetilde{\mX}^\top \widetilde{\mX}$, while part (ii) provides a technical constraint for asymptotic normality, jointly restricting the growth of $\widetilde{d}$ and the fourth moments of the transformed covariates $U^\top \widetilde{\mX}$. In the special case of i.i.d.\ errors ($U = \sigma \mI_N$), this reduces to the familiar requirement $\widetilde{d} = o(N^{1/2})$. For general $U$, the condition imposes joint restrictions on covariate moments and error dependence. When covariates are uniformly bounded, a sufficient condition is $\max_{1\leq i \leq N}\sum_{j=1}^N U_{ij} = O(1)$, which is satisfied, for example, in the linear-in-means model with row-normalized $W$.

Condition (C4) controls the truncation bias by requiring that the omitted coefficients $\|\bm{\lambda}_k\|$ decay sufficiently fast relative to the estimation error. Exponential decay arises naturally in both the linear-in-means model and its higher-order extension. We formalize this by establishing bounds on the eigenvalues of $(W^k)^\top W^k$ in the following proposition.

\begin{proposition}
\label{prop:eigen}
For any row-normalized adjacency matrix $W$ and any positive integer $k$,
$$
\lambda_{\max}\big((W^{k})^\top W^k\big) \leq N.
$$
\end{proposition}

Proposition~\ref{prop:eigen} provides a general spectral bound for arbitrary row-normalized networks. For more regular network structures, sharper bounds are available. For example, in uniform networks with bounded row and column sums, \citet{zhu2017network} show that $\lambda_{\max}\big((W^{k})^\top W^k\big) = O(\log^2 N)$. These results together yield the key inequality $a_N \leq \sqrt{N} c_{\max}^{1/2} / c_{\min}^{1/2}$, ensuring that the truncation error in Condition~(C4) decays at an appropriate rate.

In the linear-in-means model (Section~\ref{sec:introduction}), the propagation coefficients take the form $\bm{\lambda}_k = \rho^{k-1}(\rho \bm{\beta} + \bm{\delta})$ for $k \geq 1$, implying $\|\bm{\lambda}_k\| \leq |\rho|^{k} \|\bm{\beta}\| + |\rho|^{k-1} \| \bm{\delta} \|$. Assuming $\|\bm{\beta}\| \leq c d$, $\|\bm{\delta}\| \leq c d$, and $|\rho| < 1$ for some constant $c > 0$, we obtain
$$
a_N \sum_{k=K+1}^{\infty} \|\bm{\lambda}_k\| \leq \frac{2 c c_{\max}^{1/2}}{c_{\min}^{1/2} (1 - |\rho|)} \sqrt{N} d |\rho|^K.
$$
Thus, Condition~(C4) is satisfied provided $K \gg -\log(\sqrt{N}d)/\log(|\rho|)$. For instance, taking $K \geq -\log(Nd)/\log(|\rho|)$ ensures $a_N \sum_{k=K+1}^{\infty} \|\bm{\lambda}_k\| = o(1)$. Combined with Conditions~(C2) and~(C3), this indicates that $K$ should diverge at an intermediate rate: sufficiently fast to control truncation bias, but not so fast as to violate dimensionality constraints.

The preceding arguments extend naturally to higher-order variants of the linear-in-means model, although such models have not appeared in the literature, primarily due to the increased complexity of endogeneity arising from responses on both sides of the equation. Consider, for example, the following second-order linear-in-means model:
\begin{equation}
\mY = \alpha_1 \bm{1}_N + \rho_1 W \mY + \rho_2 W^2 \mY + \mX \bm{\gamma}_1 + W \mX \bm{\gamma}_2 + W^2 \mX \bm{\gamma}_3 + \mE,
\label{eq:model_higher_order}
\end{equation}
which, to the best of our knowledge, is novel. The propagation coefficients $\bm{\lambda}_k$ for $k \geq 2$ can be written (after straightforward but tedious algebra) as
\begin{align*}
\bm{\lambda}_k = & \sum_{h = \lceil k/2 \rceil}^{k} C_{h}^{k-h}  \rho_1^{2h - k} \rho_2^{k-h} \bm{\gamma}_1
+ \sum_{h = \lceil (k-1)/2 \rceil}^{k-1} C_{h}^{k-1-h} \rho_1^{2h - (k-1)} \rho_2^{k-1-h} \bm{\gamma}_2 \\
& + \sum_{h = \lceil (k-2)/2 \rceil}^{k-2} C_{h}^{k-2-h} \rho_1^{2h - (k-2)} \rho_2^{k-2-h} \bm{\gamma}_3.
\end{align*}
Assuming $|\rho_1| + |\rho_2| < 1$ and $\max\{\|\bm{\gamma}_1\|, \|\bm{\gamma}_2\|, \|\bm{\gamma}_3\|\} \leq c d$ for some $c > 0$, we obtain
$$
a_N \sum_{k=K+1}^{\infty} \|\bm{\lambda}_k\| \leq \frac{3 c c_{\max}^{1/2}}{c_{\min}^{1/2} \{1 - (|\rho_1| + |\rho_2|)^{1/2}\}^2} \sqrt{N} d \left( (|\rho_1| + |\rho_2|)^{1/2} \right)^{K-1}.
$$
Thus, Condition~(C4) is satisfied provided $K \gg -2\log(\sqrt{N}d)/\log(|\rho_1| + |\rho_2|)$; for example, $K \geq -2\log(Nd)/\log(|\rho_1| + |\rho_2|)$ suffices.

By combining Proposition~\ref{prop:eigen} with regularity conditions (C1)--(C4), we establish the asymptotic equivalence between the truncated model~\eqref{network propagation model} and the true data-generating process~\eqref{infty_model}, as formalized below.

\begin{theorem}[Estimation Consistency]
\label{thm:consistency}
Suppose Conditions (C1)--(C4) hold. Then:
\begin{enumerate}
    \item $\|\hat{\bm{\theta}} - \bm{\theta}\|^2 \xrightarrow{p} 0$, where $\bm{\theta} = (\bm{\lambda}_0^\top, \bm{\lambda}_1^\top, \ldots, \bm{\lambda}_K^\top)^\top \in \mathbb{R}^{\widetilde{d}}$ as defined in~\eqref{infty_model};
    \item $N^{-1}\|\widetilde{\mX}\hat{\bm{\theta}} - \sum_{k=0}^{\infty} W^k \mX \bm{\lambda}_k\|^2 \xrightarrow{p} 0$.
\end{enumerate}
\end{theorem}

Theorem~\ref{thm:consistency} shows that the OLS estimator $\hat{\bm{\theta}}$ from the truncated model~\eqref{network propagation model} is consistent for the true parameters in~\eqref{infty_model}. Moreover, part~(ii) establishes that the fitted values from the truncated and full models are asymptotically equivalent, justifying the use of the truncated specification for inference. To establish the asymptotic normality of $\hat{\bm{\theta}}$, we require a stronger condition:

\begin{itemize}
\item[(C5)] As $N \to \infty$ and $K \to \infty$, $\sum_{k=K+1}^{\infty} \|\bm{\lambda}_k\| = o(1/(\sqrt{N}a_N))$.
\end{itemize}

Condition (C5) strengthens Condition (C4) by requiring a faster decay of the tail coefficients, which is necessary for $\sqrt{N}$-consistency of $\hat{\bm{\theta}}$. This condition is satisfied for both the linear-in-means and second-order linear-in-means models. The main implication is that $K$ must grow slightly faster with $N$ than under Condition (C4), typically requiring a larger constant in the leading term involving $\log(Nd)$.

 Define the covariance matrix
$
\bm{\Lambda} = (N^{-1} \widetilde{\mX}^\top \widetilde{\mX}) \left(N^{-1} \widetilde{\mX}^\top U U^\top \widetilde{\mX}\right)^{-1} (N^{-1} \widetilde{\mX}^\top \widetilde{\mX}).
$
The following theorem establishes the asymptotic normality of arbitrary linear contrasts of $\hat{\bm{\theta}}$.

\begin{theorem}[Asymptotic Normality]
\label{thm:normality}
Suppose Conditions (C1)--(C5) hold. Then $\hat{\bm{\theta}}$ is asymptotically normal for the parameter vector $\bm{\theta}$ defined in~\eqref{infty_model}. Specifically, for any fixed $q < \infty$ and any $q \times \widetilde{d}$ matrix $\bm{\Psi}$ such that $\widetilde{d}^{-1}\bm{\Psi}\bm{\Psi}^\top \rightarrow \bm{G}$ for some positive-definite matrix $\bm{G} \in \mathbb{R}^{q \times q}$, as $N \to \infty$,
\begin{equation}
\widetilde{d}^{-1/2} N^{1/2} \bm{\Psi} \bm{\Lambda}^{1/2} (\hat{\bm{\theta}} - \bm{\theta}) \xrightarrow{d} N(\bm{0}_q, \bm{G}).
\end{equation}
\end{theorem}

 To conduct inference for $\bm{\theta}$ using $\hat{\bm{\theta}}$ and Theorem~\ref{thm:normality}, a consistent estimator of the noise covariance $U U^\top$ is required. Provided that the structure of $U$ is known and a consistent estimator is available, our method applies to general covariance structures. Nevertheless, to maintain theoretical clarity and avoid unnecessary complications, we restrict attention to the homoscedastic case $U = \sigma \mI_N$ in the following analysis, which corresponds to i.i.d.\ errors with variance $\sigma^2$.

Let $\mathrm{RSS} = \|\mY - \widetilde{\mX} \hat{\bm{\theta}}\|^2$ denote the residual sum of squares, and define the variance estimator $\hat{\sigma}^2 = \mathrm{RSS}/(N - \widetilde{d})$. Since model~\eqref{network propagation model} is an approximation to the true process~\eqref{infty_model}, establishing the consistency of $\hat{\sigma}^2$ for $\sigma^2$ is nontrivial. The following corollary addresses this.

\begin{corollary}[Variance Estimator Consistency]
\label{cor:consistency-variance}
Under Conditions (C1)--(C4) with $U = \sigma \mI_N$, $\hat{\sigma}^2 \xrightarrow{p} \sigma^2$.
\end{corollary}

Under the homoscedastic assumption $U = \sigma \mI_N$, consider the $j$th component $\theta_j$ of $\bm{\theta}$ for $j = 1, \ldots, \widetilde{d}$. Let $\{(\widetilde{\mX}^\top \widetilde{\mX})^{-1}\}_{jj}$ denote the $(j,j)$ entry of $(\widetilde{\mX}^\top \widetilde{\mX})^{-1}$ and define the standard error of $\hat{\theta}_j$ as
$
L_j = \hat{\sigma} \left\{(\widetilde{\mX}^\top \widetilde{\mX})^{-1}\right\}_{jj}^{1/2}.
$
By Theorem~\ref{thm:normality} and Corollary~\ref{cor:consistency-variance}, the $t$-statistic
$$
T_j = \frac{\hat{\theta}_j - \theta_j}{L_j} \xrightarrow{d} N(0, 1)
$$
as $N \to \infty$. This result provides asymptotically valid confidence intervals and hypothesis tests for individual parameter $\theta_j$.

\subsection{Determining the Relevant Neighborhood Order}
\label{sec:test}

While model~\eqref{infty_model} describes the true data-generating process, the truncated specification~\eqref{network propagation model} offers a practical and tractable approximation. Under our regularity conditions, this approximation becomes asymptotically equivalent to the full model, justifying its use for statistical inference.

Theoretically, Conditions (C1)--(C5) suggest setting $K = O(\log(Nd))$, which is sufficient for both linear-in-means and higher-order models. With this choice, we obtain the OLS estimator $\hat{\bm{\theta}}$ and turn to the important question of identifying which neighborhood orders significantly contribute to network effects.

To address this, we propose a hypothesis testing framework to determine the relevance of higher-order neighborhood information. Given the hierarchical nature of network influence, if effects vanish at a certain order, all subsequent higher-order effects should also be zero. Accordingly, we test the null hypothesis that all coefficients from order $j$ onward are zero:
\begin{itemize}
    \item[$H_0$:] $\bm{\lambda}_j = \bm{\lambda}_{j+1} = \cdots = \bm{\lambda}_K = \bm{0}$;
    \item[$H_1$:] There exists some $j' \in \{j, j+1, \ldots, K\}$ such that $\bm{\lambda}_{j'} \neq \bm{0}$.
\end{itemize}
This approach enables us to identify the maximal relevant neighborhood order, providing both statistical and substantive insight into the extent of network influence.

To implement the hypothesis test, define the restriction matrix
$$
R = \begin{bmatrix} \bm{0}_{d(K-j+1) \times dj} & \mI_{d(K-j+1)} \end{bmatrix} \in \mathbb{R}^{d(K-j+1) \times d(K+1)},
$$
where $\bm{0}_{d(K-j+1) \times dj}$ is a zero matrix that selects coefficients beyond order $j$. The Wald test statistic is then
$$
\mathcal{T}_j = N \hat{\bm{\theta}}^\top R^\top \left(R \bm{\Gamma}^{-1} R^\top\right)^{-1} R \hat{\bm{\theta}} / \hat{\sigma}^2,
$$
where $\bm{\Gamma} = N^{-1} \widetilde{\mX}^\top \widetilde{\mX}$.

When the number of restrictions $d(K-j+1)$ is fixed, classical theory gives $\mathcal{T}_j \xrightarrow{d} \chi^2(d(K-j+1))$ under $H_0$. However, when $d(K-j+1)$ grows with sample size, this approximation fails. The following theorem provides a valid asymptotic distribution in the high-dimensional regime.

\begin{theorem}[High-Dimensional Asymptotics for Wald Statistic]
\label{thm:testing}
Suppose $N^{-2} \sum_{i=1}^{N} \|\widetilde{\bm{x}}_i\|_2^4 = o(d(K-j+1))$ and $d(K-j+1) \to \infty$. Under Conditions (C1)--(C5) with $\delta_1 \geq 2$, $U = \sigma \mI_N$, and under $H_0$, as $N \to \infty$, we have
$$
\frac{\mathcal{T}_j - d(K-j+1)}{\sqrt{2d(K-j+1)}} \xrightarrow{d} N(0,1).
$$
\end{theorem}

This testing framework provides a systematic approach for identifying significant network effects by assessing whether neighborhood information of order $j$ and higher should be included in the model. In practice, we typically conduct a sequence of tests $H_{0j}: \bm{\lambda}_j = \bm{\lambda}_{j+1} = \cdots = \bm{\lambda}_K = \bm{0}$ for $j = 0, 1, \ldots, k$, where $k = o(K)$ reflects the empirical observation that second- or third-order effects are often sufficient.

For each test, the $p$-value is computed as
$$
p_j = 2\{1 - \Phi(\bZ_j)\}, \quad \text{where} \quad \bZ_j = \frac{\mathcal{T}_j - d(K-j+1)}{\sqrt{2d(K-j+1)}},
$$
and $\Phi(\cdot)$ denotes the standard normal cumulative distribution function.

Since $k$ is typically small in practice, we focus on controlling the family-wise error rate (FWER), the probability of making at least one false rejection among the multiple hypotheses. To achieve strong FWER control while maintaining reasonable power, we adopt Holm's step-down procedure \citep{holm1979simple}. This method uniformly improves upon the conservative Bonferroni correction and remains valid under arbitrary dependence among test statistics. Our simulation studies confirm that Holm's procedure effectively controls the FWER in our setting, provided $K = O(\log(Nd))$ and $k = o(K)$.

\begin{remark}
In practice, $k$ is usually small (2 or 3). If $k$ is large, Condition (C3) may be restrictive. Profiled multiple testing procedures \citep{cho2012high, lan2016testing} can address this, but are more computationally intensive and complicate interpretation. For simplicity, we focus on the current framework and leave profiled approaches for future work.
\end{remark}

\section{A Unified Framework}
\label{sec:GLM}
The truncated network propagation framework extends naturally to a wide range of response types, including binary, categorical, and time-to-event outcomes, while retaining the ability to capture multi-step network influences. We first introduce a network logistic regression model for binary data, establish its statistical properties, and highlight its connections to modern graph-based classification methods. The framework further generalizes to survival analysis via Cox's proportional hazards model, among others.

\subsection{Network Logistic Regression Model}
\label{sec:NLR}

To illustrate the versatility of our approach, we present the network logistic regression (NLR) model for binary outcomes. The NLR model incorporates multi-step network propagation by allowing each node's response to depend on covariates from neighbors up to $K$ steps away. This specification captures the influence of both direct covariates and characteristics that diffuse through the network structure.

Formally, let $Y_i \in \{0,1\}$ denote the binary outcome for node $i$. Conditional on the covariate matrix $\mX$ and the row-normalized adjacency matrix $W$, the model is specified as
\begin{equation}
P(Y_i = 1 \mid \mX, W) = P_i(\bm{\theta})
= \frac{\exp\left\{ \alpha_0 + \sum_{k=0}^{K} \bm e_i^{\top}(W^{k}\mX)\bm{\lambda}_k \right\}}
{1 + \exp\left\{ \alpha_0 + \sum_{k=0}^{K} \bm e_i^{\top}(W^{k}\mX)\bm{\lambda}_k \right\}},
\quad i = 1, \ldots, N,
\label{logit_model}
\end{equation}
where $\bm e_i$ is the $i$th standard basis vector in $\mathbb{R}^N$, and $\bm{\theta} = (\alpha_0, \bm{\lambda}_0^{\top}, \bm{\lambda}_1^{\top}, \ldots, \bm{\lambda}_K^{\top})^{\top} \in \mathbb{R}^{\tilde{d} + 1}$ collects all model parameters. Each coefficient vector $\bm{\lambda}_k$ quantifies the marginal contribution of covariates propagated through $k$ steps, thus measuring how information diffuses across successive network layers.

This extension demonstrates the broad applicability of our propagation framework. The same modeling strategy can be readily adapted to other response types, including count, categorical, and survival outcomes, providing a unified and flexible approach for analyzing network-dependent data across diverse domains.

Assuming that $Y_1, \ldots, Y_N$ are conditionally independent given $(\mX, W)$, the conditional log-likelihood is
$$
\ell(\bm{\theta} \mid \mY, \mX, W)
= \sum_{i=1}^{N} \left\{ Y_i \log P_i(\bm{\theta}) + (1 - Y_i) \log\left[1 - P_i(\bm{\theta})\right] \right\}.
$$
Maximizing $\ell(\bm{\theta} \mid \mY, \mX, W)$ yields the conditional maximum likelihood estimator (CMLE) $\widehat{\bm{\theta}}_{\mathrm{CMLE}}$.
Since a closed-form solution is generally unavailable, numerical optimization is required. When Condition (C2) holds, Newton's method provides an efficient iterative algorithm. For the network logistic regression model, the Newton-Raphson updates are given by
$$
\widehat{\bm{\theta}}^{(t+1)} = \widehat{\bm{\theta}}^{(t)} +
\left\{ \sum_{i=1}^N \omega_i(\widehat{\bm{\theta}}^{(t)}) \wt{\bm{x}}_i \wt{\bm{x}}_i^\top \right\}^{-1}
\left[ \sum_{i=1}^N \left\{Y_i - P_i(\widehat{\bm{\theta}}^{(t)})\right\} \wt{\bm{x}}_i \right],
$$
where $\omega_i(\bm{\theta}) = P_i(\bm{\theta})\{1 - P_i(\bm{\theta})\}$ is the Bernoulli variance function, and $\wt{\bm{x}}_i = \widetilde{\mX}^\top \bm{e}_i \in \mathbb{R}^{\tilde{d} + 1}$ is the extended covariate vector for node $i$, with $\widetilde{\mX} = (\bm{1}_N, \mX, W\mX, \ldots, W^K\mX)$. The algorithm iterates until numerical convergence.

For notational clarity, let $x_{ij}$ and $\wt{x}_{ij}$ denote the $(i,j)$-th elements of $\mX$ and $\wt{\mX}$, respectively.

\begin{itemize}
\item[(C3')] The covariates are uniformly bounded: $\max_{i,j} |x_{ij}| < \infty$.
\end{itemize}

Condition (C3') guarantees that all covariate values are bounded. Since $W$ is row-normalized, each power $W^k$ also remains row-normalized, ensuring that the elements of the extended design matrix $\widetilde{\mX} = (\bm{1}_N, \mX, W\mX, \ldots, W^K\mX)$ satisfy
$$
\max_{i,l} |\widetilde{x}_{il}| \leq \max\left\{1, \max_{m,j} |x_{mj}|\right\} < \infty.
$$
This boundedness facilitates the derivation of asymptotic normality for the CMLE by leveraging established results for logistic regression with diverging dimensionality and fixed design \citep{Liang2012}. Define the information matrix $\bm{\Gamma}_N = N^{-1}\sum_{i=1}^N \omega_i(\bm{\theta}) \widetilde{\bm{x}}_i \widetilde{\bm{x}}_i^\top$, where $\omega_i(\bm{\theta}) = P_i(\bm{\theta})\{1 - P_i(\bm{\theta})\}$.

\begin{theorem}[Asymptotic Normality for Binary Responses]
\label{thm:normal-binary}
Under Conditions (C2), (C3)(i), and (C3'), for any fixed $q < \infty$ and any sequence of $q \times \widetilde{d}$ matrices $\bm{\Psi}$ satisfying $\widetilde{d}^{-1}\bm{\Psi}\bm{\Psi}^\top \to \bm{G}$ for some nonnegative definite symmetric matrix $\bm{G} \in \mathbb{R}^{q \times q}$, we have, as $N \to \infty$,
$$
\widetilde{d}^{-1/2} N^{1/2} \bm{\Psi} \bm{\Gamma}_N^{1/2} (\widehat{\bm{\theta}}_{\text{CMLE}} - \bm{\theta}) \xrightarrow{d} N(\bm{0}_q, \bm{G}).
$$
\end{theorem}

The network logistic regression model~\eqref{logit_model} extends standard logistic regression by incorporating multi-step network effects through propagated covariates $W^k\mX$. This approach preserves computational tractability, yields interpretable coefficients that distinguish direct and indirect effects, and naturally generalizes classical models as a special case when $K=0$.

To highlight the computational simplicity of our NLR framework, we compare it to the classical spatial probit model, which generalizes the linear-in-means model to binary outcomes via a latent utility formulation:
\begin{align}
\mZ &= \rho W \mZ + \mX \bm{\gamma} + \mE, \quad \mE \sim \mathcal{N}(\bm{0}_N, \mI_N), \\
Y_i &= \mathbf{1}\{Z_i > 0\}, \quad i = 1, \dots, N.
\end{align}
Inference for the spatial probit model requires evaluating $N$-dimensional multivariate normal probabilities, which is computationally intensive ($O(N^3)$) and quickly becomes infeasible for large networks. In addition, efficient estimation often imposes restrictive assumptions on network structure \citep{lee2004asymptotic}.

By contrast, our NLR model operates directly on the observed likelihood, avoiding latent variables and high-dimensional integration. This leads to scalable estimation and practical applicability for large networks. As shown in Section~5, NLR achieves superior computational efficiency and predictive performance compared to both traditional logistic regression and deep learning approaches, while maintaining interpretability.

\subsection{Network Cox Proportional Hazards Model}
\label{sec:Cox_model}

Our propagation framework extends naturally to survival analysis, enabling principled modeling of time-to-event data with censoring in networked settings. Let $T_i$ and $C_i$ denote the failure and censoring times for node $i$, with observed time $\tilde{T}_i = \min\{T_i, C_i\}$ and event indicator $\delta_i = \mathbf{1}\{T_i \leq C_i\}$. Assuming conditional independence of $T_i$ and $C_i$ and non-informative censoring given $(\mX, W)$, we define the network Cox model with hazard function
\begin{equation}
h(t \mid \mX, W) = h_0(t) \exp\left( \sum_{k=0}^{K} \bm{e}_i^\top (W^k \mX) \bm{\lambda}_k \right),
\end{equation}
where $h_0(t)$ is an unspecified baseline hazard and $\bm{\lambda}_k$ are the propagation coefficients. The inclusion of $W^k \mX$ allows the model to capture both direct and higher-order network effects on the hazard.
The corresponding log partial likelihood is
\begin{equation}
\ell_N(\bm{\lambda}) = \sum_{i=1}^N \delta_i \left[ \sum_{k=0}^{K} \bm{e}_i^\top (W^k \mX) \bm{\lambda}_k - \log\left( \sum_{j: \tilde{T}_j \geq \tilde{T}_i} \exp\left\{ \sum_{k=0}^{K} \bm{e}_j^\top (W^k \mX) \bm{\lambda}_k \right\} \right) \right],
\label{eq:cox_partial_likelihood}
\end{equation}
where $\bm{\lambda} = (\bm{\lambda}_0^\top, \bm{\lambda}_1^\top, \ldots, \bm{\lambda}_K^\top)^\top \in \mathbb{R}^{\tilde{d}}$. The maximum partial likelihood estimator $\widehat{\bm{\lambda}} = \arg \max_{\bm{\lambda}} \ell_N(\bm{\lambda})$ is consistent and efficient under standard regularity conditions.

This formulation generalizes the classical Cox model \citep{cox1972regression, cox1975partial}: for $K = 0$, it reduces to standard Cox regression, while for $K > 0$, it incorporates interpretable network diffusion effects. Thus, our approach provides a unified and flexible framework for survival analysis on networks.

Following \citet{andersen1982cox}, we employ counting process theory to establish the asymptotic properties of $\widehat{\bm{\lambda}}$. Let $N_i(t) = \mathbf{1}\{T_i \leq t, T_i \leq C_i\}$ denote the counting process and $Y_i(t) = \mathbf{1}\{\tilde{T}_i \geq t\}$ the at-risk process for node $i$. To accommodate time-dependent covariates, we allow $\mX$ to vary with $t$, denoted $\mX(t)$, with $x_{ij}(t)$ as its $(i,j)$-th entry. Without loss of generality, we assume $t \in [0,1]$. We impose the following regularity conditions:
\begin{itemize}
    \item[(C5)] The cumulative baseline hazard is uniformly bounded: $\sup_{t \in [0,1]} h_0(t) \leq C$ for some $C < \infty$.
    \item[(C6)] The processes $\mX(t)$ and $Y_i(t)$ are left-continuous with right-hand limits. There exist constants $C_1 < \infty$ and $C_2 > 0$ such that $\max_{t \in [0,1]} \max_{ij} |x_{ij}(t)| \leq C_1$ and $\min_{i} P\{Y_i(t) = 1, \ \forall t \in [0,1]\} > C_2$.
    \item[(C7)] For the true parameter $\bm{\lambda}^*$, there exists a neighborhood $\mathcal{B}$ and constant $C_B < \infty$ such that $\max_{t \in [0,1]} \max_{i} \max_{\bm{\lambda} \in \mathcal{B}} |\widetilde{\bm{x}}_i(t)^\top \bm{\lambda}| \leq C_B$, where $\widetilde{\bm{x}}_i(t)^\top = \bm{e}_i^\top (\mX(t), W\mX(t), \ldots, W^K\mX(t))$.
    \item[(C8)] Define $s^{(k)}(\bm{\lambda}, t) = \mathbb{E}[N^{-1} \sum_{i=1}^N \widetilde{\bm{x}}_i(t)^{\otimes k} Y_i(t) \exp\{\widetilde{\bm{x}}_i(t)^\top \bm{\lambda}\}]$ for $k = 0,1,2$, with $s^{(0)}$ bounded away from zero. These functions are uniformly continuous and bounded on $\mathcal{B} \times [0,1]$.
    \item[(C9)] The Fisher information matrix
    $
    \bm{V}_N(\bm{\lambda}) = \int_0^1 \bm{v}(\bm{\lambda}, t) s^{(0)}(\bm{\lambda}, t) h_0(t) dt,
    $
    where
    $$
    \bm{v}(\bm{\lambda}, t) = \frac{s^{(2)}(\bm{\lambda}, t)}{s^{(0)}(\bm{\lambda}, t)} - \left\{ \frac{s^{(1)}(\bm{\lambda}, t)}{s^{(0)}(\bm{\lambda}, t)} \right\} \left\{ \frac{s^{(1)}(\bm{\lambda}, t)}{s^{(0)}(\bm{\lambda}, t)} \right\}^\top,
    $$
    satisfies $\|\bm{V}_N(\bm{\lambda}^*) - \bm{V}_\infty\|_2 \to 0$ for some positive definite $\bm{V}_\infty$ as $N \to \infty$.
\end{itemize}

Conditions (C5)--(C9) collectively guarantee the consistency of $\widehat{\bm{\lambda}}$ and the asymptotic normality of its linear contrasts. Specifically, (C5) bounds the cumulative hazard, while (C6) ensures the applicability of functional laws of large numbers in $D[0,1]$. Condition (C7) controls the moments of the weighted covariates, and, together with (C6) and (C8), establishes the uniform convergence of the empirical processes $S^{(k)}(\bm{\lambda}, t)  = N^{-1} \sum_{i=1}^N \widetilde{\bm{x}}_i(t)^{\otimes k} Y_i(t) \exp\{\widetilde{\bm{x}}_i(t)^\top \bm{\lambda}\}$ to their population limits \citep{andersen1982cox, fan2002variable}. Condition (C9) strengthens the standard positive semi-definiteness requirement to guarantee identifiability. Analogous conditions are standard in the survival analysis literature \citep{andersen1982cox, fan2002variable, zhang2007adaptive}.

\begin{theorem}[Asymptotic Normality for Network Cox Model]
\label{thm:cox_asymptotics}
Suppose that, conditional on $W$ and $\mX$, the pairs $(T_i, C_i)$ are independent for $i = 1, \dots, N$, and the parameter space is compact. Under Conditions (C2) with $\alpha_1 < 1/2$ and (C5)--(C9), the maximum partial likelihood estimator satisfies
$$
N^{1/2} \bm{u}^\top (\widehat{\bm{\lambda}} - \bm{\lambda}^*) \xrightarrow{d} N\left(0, \bm{u}^\top \bm{V}_\infty^{-1} \bm{u}\right)
$$
for any fixed unit vector $\bm{u}$, where $\bm{V}_\infty$ is as defined in Condition (C9).
\end{theorem}

This theorem establishes the $\sqrt{N}$-consistency and asymptotic normality of the network Cox model estimator, confirming that the propagation framework retains optimal statistical properties in the presence of time-dependent covariates and censoring. The asymptotic variance $\bm{V}_\infty^{-1}$ attains the semiparametric efficiency bound, ensuring optimal inference. Consequently, standard procedures such as Wald tests and confidence intervals for the network diffusion parameters $\bm{\lambda}_k$ are valid, facilitating rigorous application to survival data with network dependence.

\section{Simulation}
\label{sec:simulation}
We conduct extensive simulation studies to rigorously evaluate the performance of our proposed methodology. The experiments are designed to assess the ability of the truncated network propagation model~\eqref{network propagation model} to approximate diverse data-generating processes, with particular emphasis on scenarios involving model misspecification.

\subsection{Data Generation Mechanisms}
\label{sec:simulation1}
To benchmark our approach under varying network and covariate structures, we consider three  cases, corresponding to Cases 1–3 below.
%In all cases, the intercept is set to zero and the covariate dimension is fixed at $d = 10$. Error terms $\ve_i$ are independently drawn from $N(0,1)$.
The network adjacency matrix $A$ and covariate matrix $\mX$ are specified as follows:

\noindent
\textbf{Case 1: Erd\H{o}s-R\'enyi Random Graph}
We first consider the classical Erd\H{o}s-R\'enyi model \citep{Erdos:Renyi:1959}, where edges are formed independently with probability $P(a_{ij}=1) = N^{-0.8}$. Network sizes $N \in \{1000, 1500, 2000\}$ yield densities from 0.4\% to 0.23\%. Covariates $X_i = (X_{i1}, \dots, X_{id})^\top$ are generated from a multivariate normal distribution with mean zero and covariance $\Sigma_X$ defined by $(\Sigma_X)_{j_1j_2} = 0.5^{|j_1-j_2|}$.

\noindent
\textbf{Case 2: Stochastic Block Model}
To evaluate performance in the presence of community structure, we employ the stochastic block model \citep{Nowicki:Snijders:2001}. Nodes are assigned to one of three blocks with equal probability. Within-block edges are formed with probability $N^{-0.75}$, and between-block edges with probability $N^{-1}$. We consider $N \in \{1000, 2000, 2500\}$, with densities decreasing from 0.4\% to 0.2\%. Covariates are generated as in Case 1.

\noindent
\textbf{Case 3: Power-Law Network}
To mimic real-world networks with heavy-tailed degree distributions \citep{adamic2000power}, in-degrees $|\mathcal{N}_i|$ are sampled from a discrete power-law distribution $P(|\mathcal{N}_i| = k) = c k^{-2.5}$, where $c$ is a normalizing constant. Each node randomly selects $|\mathcal{N}_i|$ followers. Network sizes $N \in \{1000, 1500, 2000\}$ yield densities from 0.29\% to 0.15\%. Covariates are multivariate normal with $(\Sigma_X)_{j_1j_2} = 0.5$ for $1 \leq j_1 \neq j_2 \leq d-1$, $(\Sigma_X)_{j_1j_2} = \sqrt{0.5}$ if $j_1 = d$ or $j_2 = d$, and $(\Sigma_X)_{jj} = 1$ for all $j$.

We generate response variables according to four distinct data-generating mechanisms (Settings 1--4, described below) to assess the flexibility and robustness of model~\eqref{network propagation model} under a range of structural assumptions. The covariate dimension is fixed at $d = 10$, and the error terms $\ve_i$ are independently drawn from $N(0,1)$. In Settings 1--3, the intercept is set to zero.

\noindent
\textbf{Setting 1: Linear-in-Means Model}
Responses are generated from the classical linear-in-means model \eqref{linear in mean model} with autoregressive parameter $\rho = 0.25$. Regression coefficients $\bm{\beta}, \bm{\delta} \in \mathbb{R}^d$ are independently drawn from $\text{Uniform}(0.5, 5)$. The  network structure $A$ and covariate $\mX$ are generated as described in Cases 1--3, and the response vector $\mY$ is computed accordingly.

\noindent
\textbf{Setting 2: Higher-Order Network Influence}
We consider a higher-order linear-in-means model~\eqref{eq:model_higher_order} with parameters $\rho_1 = 0.25$ and $\rho_2 = 0.05$. Coefficient vectors $\bm{\gamma}_1, \bm{\gamma}_2, \bm{\gamma}_3 \in \mathbb{R}^d$ are independently sampled from $\text{Uniform}(0.5, 5)$. The response vector $\mY$ is generated for each network configuration in Cases 1--3.

\noindent
\textbf{Setting 3: Network Propagation Model}
Data are simulated from the network propagation model~\eqref{infty_model}, with coefficients $\bm{\lambda}_k \in \mathbb{R}^d$ for $k = 0, \ldots, 5$ independently drawn from $\text{Uniform}(0, 5)$, and $\bm{\lambda}_k = 0$ for $k \geq 6$. This setting directly evaluates our proposed framework under its assumed generative process.

\noindent
\textbf{Setting 4: Linear Regression with Network Cohesion}
We implement the network cohesion model~\eqref{RNC} introduced by \citet{li2019prediction}:
$
\mY = \bm{\mu} + \mX \bm{\beta} + \mE,
$
where $\bm{\mu} = (\mu_1, \ldots, \mu_N)^\top$ represents node-specific effects. Community memberships $c_i \in \{1, 2, 3\}$ are assigned uniformly at random. For each node, $\mu_i \sim N(\eta_{c_i}, 0.25)$, where the community means are set to $\eta_1 = -2.5$, $\eta_2 = 0$, and $\eta_3 = 2.5$. Regression coefficients are independently drawn as $\beta_j \sim N(1, 1)$ for $j = 1, \ldots, d$.

Model~\eqref{RNC} is tailored for networks with pronounced community structure, promoting similar predictions among nodes within the same community. Accordingly, for this setting, we exclusively use Case 2 to generate $A$ and $\mX$, and simulate $\mY$ as specified above.

\subsection{Performance Evaluation}
\label{sec:validation}
%Extensive simulations are conducted  to assess the empirical performance of the truncated network propagation model~\eqref{network propagation model}.
For each data-generating scenario, 100 independent datasets are generated, and results are averaged to ensure statistical robustness. Our primary goal is to benchmark predictive accuracy of model ~\eqref{network propagation model} against established alternatives, including the linear-in-means model, its higher-order extension~\eqref{eq:model_higher_order}, and the network cohesion model~\eqref{RNC}.

Since the true order of network dependence is typically unknown in practice, we fix the truncation parameter at $K = 8$, motivated by its asymptotic order $O(\log(Nd))$. To mitigate multicollinearity from multiple network lags, we apply a forward selection procedure to extract linearly independent columns from the extended design matrix $(\mX, W\mX, \ldots, W^K\mX)$.

Model performance is evaluated via prediction errors in four scenarios: (1) oracle performance using the true model; (2) in-sample fit; (3) out-of-sample prediction with network connections between training and test data; and (4) out-of-sample prediction with isolated test networks. The distinction between (3) and (4) is particularly relevant for network data, as real-world applications may or may not feature test nodes connected to the training set. Precise mathematical definitions of all prediction error metrics are provided in Appendix~B. For scenario (3), each dataset is randomly split into 80\% training and 20\% testing. For scenario (4), an independent test set of equal size as that in scenario (3) is generated to ensure isolation.

Competing models are estimated using efficient GMM~\citep{LEE2014}, which yields consistent estimates without the computational overhead of quasi-maximum likelihood. Comparative performance is summarized using relative ratios: training performance versus oracle ($\kappa_1$), in-sample performance ($\kappa_2$), and out-of-sample performance for both network scenarios ($\kappa_3$, $\kappa_4$). Ratios below 1 indicate superior performance of our method, with smaller values reflecting greater advantage.

Table~\ref{tab:1} reports the performance of the proposed network propagation model~\eqref{network propagation model} when applied to data generated from the linear-in-means model (Setting 1) and its higher-order extension (Setting 2).
Table~\ref{tab:2} presents results for the complementary scenario, where data are generated from the network propagation model~\eqref{infty_model} (Setting 3) and fitted using competing models.
\begin{table}[!h]
\renewcommand\arraystretch{1.5}
\centering
\caption{Comparative performance of the network propagation model against true data-generating processes. First block: Data are generated from Setting 1 (linear-in-means); Second block:  Data are generated from Setting 2 (higher-order linear-in-means). Values below 1 indicate superior performance of our method.}
\label{tab:1}
\vspace{0.25cm}
{\scriptsize
\begin{tabular}{ccccccccccccccccc}
\hline
& \multicolumn{5}{c}{Case 1} && \multicolumn{4}{c}{Case 2} && \multicolumn{4}{c}{Case 3} \\
\cmidrule{1-5} \cmidrule{7-11} \cmidrule{13-17}
N & $\kappa_1$ & $\kappa_2$ & $\kappa_3$ & $\kappa_4$ && N & $\kappa_1$ & $\kappa_2$ & $\kappa_3$ & $\kappa_4$ && N & $\kappa_1$ & $\kappa_2$ & $\kappa_3$ & $\kappa_4$ \\
\hline
\multicolumn{16}{c}{\textit{Model \eqref{network propagation model} vs. Linear-in-Means Model \eqref{linear in mean model}}} \\
\hline
1000&0.969& 0.988&1.049&1.228&&
1000&0.969&0.989&1.047 &      1.173
&&1000& 0.969&0.989&1.065&1.071   \\
1500&  0.981 &      0.994 &      1.032 &      1.098 &&  2000&  0.985 &      0.996 &      1.022 &      1.084 &&   1500& 0.978 &      0.991 &      1.052 &      1.035   \\
2000&  0.985 &      0.995 &      1.022 &      1.078 &&  2500&  0.990 &      0.998 &      1.017 &      1.057 &&   2000& 0.983 &      0.994 &      1.031 &      1.027   \\
\hline
\multicolumn{16}{c}{\textit{Model \eqref{network propagation model} vs. Higher-Order Model \eqref{eq:model_higher_order}}} \\
\hline
1000&   0.966 &      0.994 &      1.058 &      1.240 &&  1000&  0.965 &      0.992 &      1.064 &      1.265 &&  1000&  0.958 &      0.981 &      1.070 &      1.066  \\
1500&   0.977 &      0.996 &      1.036 &      1.148 &&  2000&  0.985 &      1.000 &      1.028 &      1.175 &&  1500&  0.971 &      0.985 &      1.041 &      1.046   \\
2000&   0.984 &      0.998 &      1.031 &      1.088 &&  2500&  0.988 &      1.000 &      1.023 &      1.099 &&  2000&  0.978 &      0.989 &      1.030 &      1.030  \\

\hline
\end{tabular}}
\end{table}

\begin{table}[!h]
\renewcommand\arraystretch{1.5}
\centering
\caption{Performance comparison when data are generated from the network propagation model (Setting 3). First block:  model \eqref{network propagation model} vs. linear-in-means model; Second block: model \eqref{network propagation model} vs. higher-order linear-in-means model.}
\label{tab:2}
\vspace{0.25cm}
{\scriptsize
\begin{tabular}{ccccccccccccccccc}
\hline
& \multicolumn{5}{c}{Case 1} && \multicolumn{4}{c}{Case 2} && \multicolumn{4}{c}{Case 3} \\
\cline{1-5} \cline{7-11} \cline{13-17}
N & $\kappa_1$ & $\kappa_2$ & $\kappa_3$ & $\kappa_4$ && N & $\kappa_1$ & $\kappa_2$ & $\kappa_3$ & $\kappa_4$ && N & $\kappa_1$ & $\kappa_2$ & $\kappa_3$ & $\kappa_4$ \\
\hline
\multicolumn{16}{c}{\textit{ Model \eqref{network propagation model} vs.  Linear-in-Means Model \eqref{linear in mean model}  }} \\
\hline
1000&   0.955 &   0.101 &  0.117 &   0.049 &&   1000&  0.957 &   0.109 &  0.123 &   0.051 &&  1000&  0.944 &   0.006 &  0.007 &   0.007 \\
1500&   0.970 &   0.128 &  0.139 &   0.055 &&   2000&  0.981 &   0.181 &  0.193 &   0.064 &&  1500&  0.964 &   0.006 &  0.007 &   0.007  \\
2000&   0.979 &   0.155 &  0.163 &   0.062 &&   2500&  0.984 &   0.209 &  0.217 &   0.074 &&  2000&  0.973 &   0.006 &  0.006 &   0.006  \\
\hline
\multicolumn{16}{c}{\textit{  Model \eqref{network propagation model} vs. Higher-Order  Model \eqref{eq:model_higher_order} }} \\
\hline
1000&   0.955 &   0.225 &  0.245 &   0.084 &&   1000&  0.957 &   0.231 &  0.252 &   0.090 &&   1000&  0.944 &   0.012 &  0.014 &   0.009  \\
1500&   0.970 &   0.309 &  0.328 &   0.093 &&   2000&  0.981 &   0.412 &  0.411 &   0.130 &&   1500&  0.964 &   0.014 &  0.014 &   0.012  \\
2000&   0.979 &   0.377 &  0.381 &   0.140 &&   2500&  0.984 &   0.478 &  0.486 &   0.161 &&   2000&  0.973 &   0.015 &  0.015 &   0.012\\
\hline
\end{tabular}}
\end{table}

The results in Table~\ref{tab:1} highlight the strong approximation capability of the truncated network propagation model~\eqref{network propagation model}. As $N$ increases, all performance ratios approach 1, indicating that our model effectively recovers both the linear-in-means specification and its higher-order extensions. The consistent proximity of $\kappa_1$ to 1 across scenarios further demonstrates that our approach closely approximates the true data-generating process. While mild overfitting is observed in finite samples (with $\kappa_1$ and $\kappa_2$ slightly below 1), this effect diminishes as network size grows, underscoring the asymptotic consistency of our method.

In contrast, Table~\ref{tab:2} reveals that when the true data-generating process follows the network propagation model, traditional linear-in-means models exhibit substantial performance loss, as reflected by the markedly lower $\kappa_2$--$\kappa_4$ values. This underscores the robustness and flexibility of our framework, particularly under model misspecification.

Finally, it is noteworthy that the values of $\kappa_4$ consistently approach 1 in Table~\ref{tab:1}, while remaining substantially below 1 in Table~\ref{tab:2}. This pattern demonstrates that our model maintains strong predictive performance even when test networks are isolated from the training data. In contrast, the more variable $\kappa_3$ values highlight the added benefit of leveraging cross-set network connections when they are available.

\begin{figure}[htbp]
    \centering
    \begin{minipage}{0.48\textwidth}
        \centering
        \includegraphics[width=\linewidth]{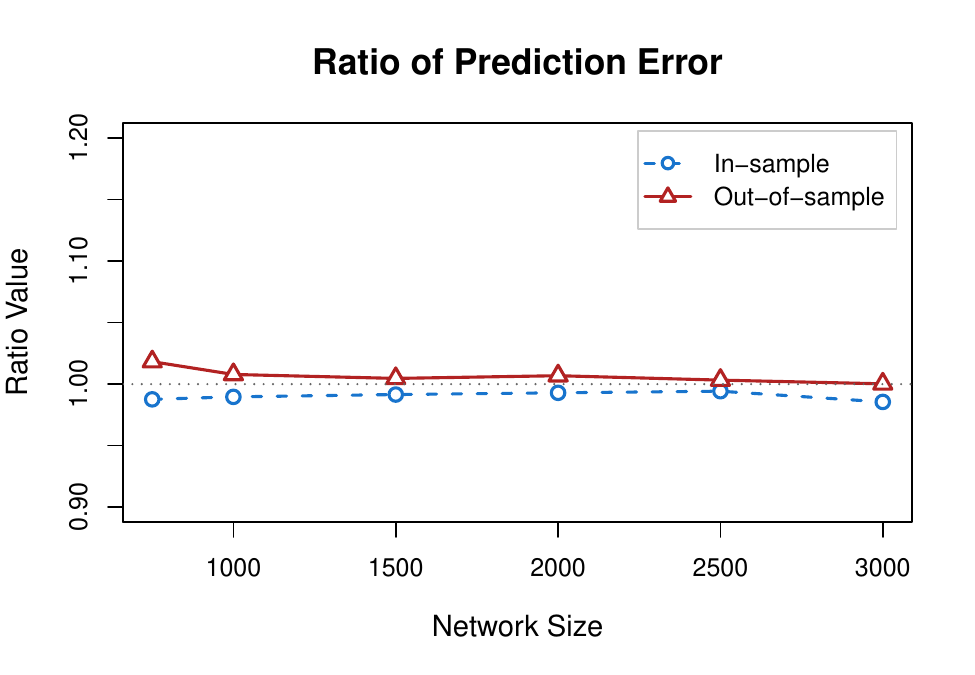}
    \end{minipage}
    \hfill
    \begin{minipage}{0.48\textwidth}
        \centering
        \includegraphics[width=\linewidth]{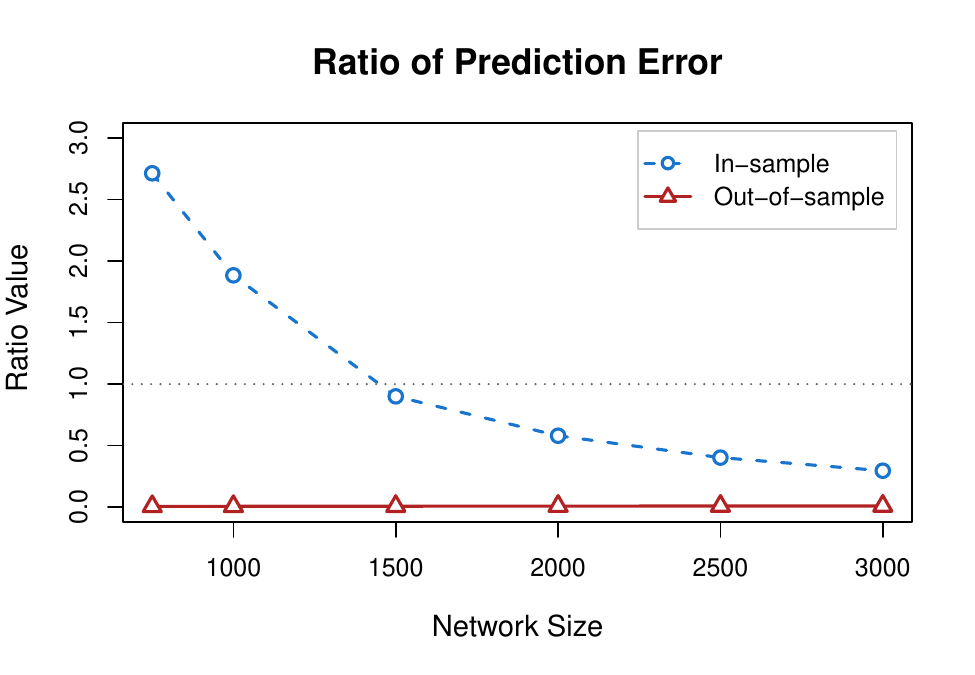}
    \end{minipage}
    \caption{
        Robustness comparison between the network propagation and cohesion models.
        \textbf{Left:} Data generated from the network cohesion model (Setting~4). Our method achieves competitive performance, demonstrating robustness to model misspecification.
        \textbf{Right:} Data generated from the network propagation model (Setting~3). Our method consistently outperforms the cohesion model, which exhibits overfitting for small network sizes.
    }
    \label{fig1}
\end{figure}

We compare the proposed network propagation model~\eqref{network propagation model} with the network cohesion model~\eqref{RNC} under Settings~3 and~4. For this focused comparison, networks are generated using the Stochastic Block Model (Case~2), and performance is evaluated across network sizes $N \in [750, 3000]$. Following \citet{li2019prediction}, we report both in-sample ($\kappa_2$) and out-of-sample ($\kappa_3$) prediction error ratios, with dashed blue lines representing $\kappa_2$ and solid red lines representing $\kappa_3$ in Figure~\ref{fig1}.

Figure~\ref{fig1} illustrates the contrasting performance patterns under different data-generating mechanisms. The left panel shows that when data are generated from the network cohesion model (Setting~4), our method remains competitive, with both $\kappa_2$ and $\kappa_3$ converging to 1 as $N$ increases. This convergence highlights the robustness of our framework to model misspecification, achieving accuracy comparable to the true model despite the structural mismatch.

In contrast, the right panel demonstrates that when data are generated from the network propagation model (Setting~3), our method consistently outperforms the cohesion model (values below 1), while the cohesion model exhibits significant overfitting ($\kappa_2 > 1$) {for smaller network sizes} and poor generalization ($\kappa_3 < 1$). This overfitting is attributable to the cohesion model's large parameter space, which increases both statistical variability and computational burden.

Collectively, these results highlight the superior flexibility of our network propagation framework. Unlike specialized methods that perform optimally only under specific data-generating processes, our approach consistently achieves robust performance across a variety of scenarios. This demonstrates its practical utility for real-world applications, where the true structure of network influence is often unknown.

In summary, the truncated network propagation model~\eqref{network propagation model} offers three key advantages: (1) asymptotic convergence to the true data-generating process as $N$ increases; (2) robust performance across diverse network structures and generative mechanisms; and (3) effective modeling of higher-order neighborhood influences. In contrast, both linear-in-means models and the network cohesion approach exhibit sensitivity to model specification, with their performance highly dependent on correct model identification.

\subsection{Finite-Sample Performance of Hypothesis Tests}
\label{sec:simu_test}

We assess the finite-sample performance of our proposed testing procedure by evaluating $k = 5$ sequential hypotheses. Let $\mathcal{N}_0 = \{j: \bm{\lambda}_j = \bm{\lambda}_{j+1} = \cdots = \bm{\lambda}_K = \bm{0},\, 0 \leq j < k\}$ denote the set of true null hypotheses, and $\mathcal{N}_1 = \{j: \bm{\lambda}_j \neq \bm{0},\, 0 \leq j < k\}$ the set of false nulls. We consider two scenarios: $|\mathcal{N}_0| = 2$ (both first- and second-order neighborhood effects present) and $|\mathcal{N}_0| = 3$ (only first-order effects), representing different levels of sparsity in network effects.

Following Section~\ref{sec:simulation1}, we fix the truncation parameter at $K = 8$ when fitting model~\eqref{network propagation model}. Simulations are conducted across three network structures (Cases 1--3) and varying network sizes. For $|\mathcal{N}_0| = 2$, we generate $\bm{\lambda}_j \in \mathbb{R}^d$ for $j = 0, 1, 2$ with entries independently drawn from $\text{Uniform}(-0.25, 0.25)/\sqrt{2}$; for $|\mathcal{N}_0| = 3$, the same applies for $j = 0, 1$. Data are generated according to model~\eqref{infty_model}, with each scenario replicated 1,000 times at significance level $\xi = 0.05$.

We evaluate the testing procedure's ability to detect network effects at different distances. For each effect distance $j$, we test whether effects beyond that distance are zero. Performance is assessed using four key metrics, averaged over 1,000 simulations: a) Type I Error Rate: Frequency of false positives,
    b) Power: Frequency of correctly detecting true effects,
   c) Multiple Testing Control: Family-wise error rates when testing multiple distances, and
    d) Coverage: Empirical coverage of confidence intervals for estimated effects.

Table~\ref{tab:3} summarizes results across network sizes and structures. Our testing procedure demonstrates strong finite-sample properties: type I error rates approach the nominal 5\% level as network size increases, power converges to near-perfect detection, multiple testing adjustments effectively control family-wise error rates without substantial power loss, and confidence intervals maintain approximately 95\% coverage. These findings indicate that our framework provides reliable statistical inference for network effects, with appropriate error control and high detection power across practical sample sizes.

\begin{table}[!h]
\renewcommand\arraystretch{1.5}
\centering
\setlength\tabcolsep{5pt}
\caption{Testing performance across network sizes and structures. Columns show: empirical power for true effects (EP), type I error rate (ES), multiple testing power (MP), family-wise error rate (FWER), and confidence interval coverage (CP). Target values: EP = 1.0, ES = 0.05, CP = 0.95.}
\label{tab:3}
\vspace{0.25cm}
{\scriptsize
\begin{tabular}{cccccccccccccccccc}
\hline
& \multicolumn{5}{c}{Case 1} && \multicolumn{5}{c}{Case 2} && \multicolumn{5}{c}{Case 3} \\
\cmidrule{2-6} \cmidrule{8-12} \cmidrule{14-18}
$N$ & EP & ES & MP & FWER & CP && EP & ES & MP & FWER & CP && EP & ES & MP & FWER & CP \\
\hline
\multicolumn{18}{c}{\textit{2 True Null Hypotheses}} \\
\hline
1000 & 0.684 & 0.086 & 0.622 & 0.023 & 0.949 && 0.676 & 0.100 & 0.601 & 0.028 & 0.951 && 0.815 & 0.109 & 0.764 & 0.035 & 0.952 \\
2000 & 0.737 & 0.060 & 0.680 & 0.014 & 0.951 && 0.719 & 0.061 & 0.672 & 0.014 & 0.952 && 0.922 & 0.066 & 0.880 & 0.018 & 0.950 \\
3000 & 0.755 & 0.053 & 0.721 & 0.011 & 0.950 && 0.738 & 0.055 & 0.700 & 0.015 & 0.952 && 0.981 & 0.056 & 0.962 & 0.013 & 0.952 \\
\hline
\multicolumn{18}{c}{\textit{3 True Null Hypotheses}} \\
\hline
1000 & 0.811 & 0.091 & 0.736 & 0.037 & 0.948 && 0.823 & 0.117 & 0.746 & 0.048 & 0.949 && 0.908 & 0.121 & 0.857 & 0.051 & 0.953 \\
3000 & 0.982 & 0.067 & 0.959 & 0.019 & 0.950 && 0.970 & 0.064 & 0.940 & 0.019 & 0.948 && 1.000 & 0.069 & 1.000 & 0.023 & 0.948 \\
5000 & 1.000 & 0.052 & 0.997 & 0.018 & 0.950 && 0.998 & 0.053 & 0.994 & 0.016 & 0.949 && 1.000 & 0.049 & 1.000 & 0.018 & 0.949 \\
\hline
\end{tabular}}
\end{table}

\section{Social Media Sentiment Analysis}
\label{sec:real_data}

We illustrate the practical utility of our methodology with a sentiment analysis of social media responses to the \emph{``Celebrity Civil-Service Examination Incident''} in July 2022. This event, in which three Chinese celebrities applied for civil-service positions, sparked widespread public discussion and provides an ideal setting for studying network-dependent emotional contagion.

\subsection{Data Collection and Processing}

User comments and follower networks were collected from Sina Weibo (\url{http://www.weibo.com}) using a snowball sampling strategy; detailed procedures are provided in Appendix~C. The final dataset consists of $N = 8,101$ users with complete network and textual information.

Sentiment labels were assigned through rigorous manual annotation: each comment was independently evaluated by at least two professional annotators. After excluding neutral samples, we set $Y = 1$ for positive sentiment and $Y = 0$ for negative sentiment, ensuring high inter-annotator reliability. Following established Chinese emotion classification frameworks, we extracted seven emotion categories from the comment texts. For each user, we computed word frequency scores for these categories, generating predictor variables that capture the multidimensional nature of emotional expression.

\begin{figure}[!htbp]
    \centering
    \begin{subfigure}[t]{0.49\textwidth}
        \centering
        \includegraphics[width=\textwidth]{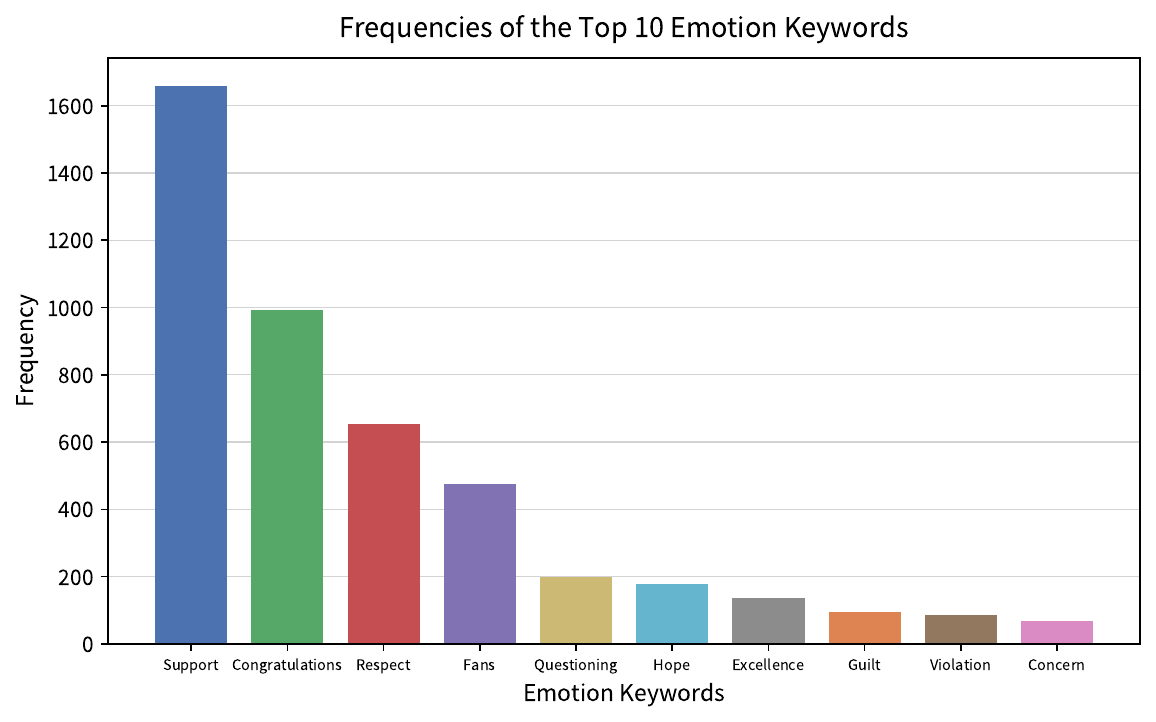}
    \end{subfigure}
    \hfill
    \begin{subfigure}[t]{0.49\textwidth}
        \centering
        \includegraphics[width=\textwidth]{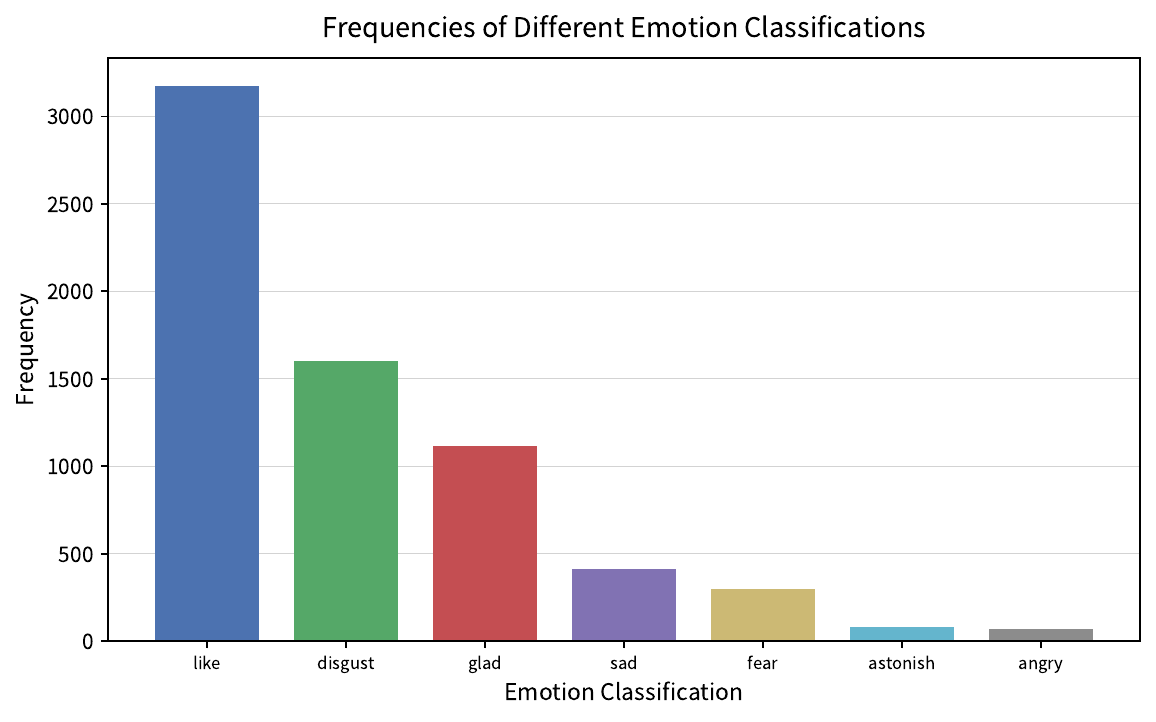}
    \end{subfigure}
    \caption{Frequencies of emotion-related keywords (left) and emotion categories (right) in user comments.}
    \label{fig:emotion-frequency}
\end{figure}

We begin with preliminary descriptive analyses to characterize the emotional landscape of user responses. First, we compute the frequency of each emotion-related keyword in user comments and present the ten most common keywords in the left panel of Figure~\ref{fig:emotion-frequency}. The right panel summarizes the distribution of emotion categories, where each category is counted once per observation if it appears at least once.

Figure~\ref{fig:emotion-frequency} reveals that a substantial proportion of users express supportive attitudes toward the celebrities' applications, with many positive comments likely originating from fans. However, the prevalence of certain high-frequency keywords, together with the distribution of emotion categories, also indicates a notable presence of opposition or negative sentiment regarding the incident.

\begin{figure}[!htbp]
    \centering
    \begin{subfigure}[t]{0.49\textwidth}
        \centering
        \includegraphics[width=\textwidth]{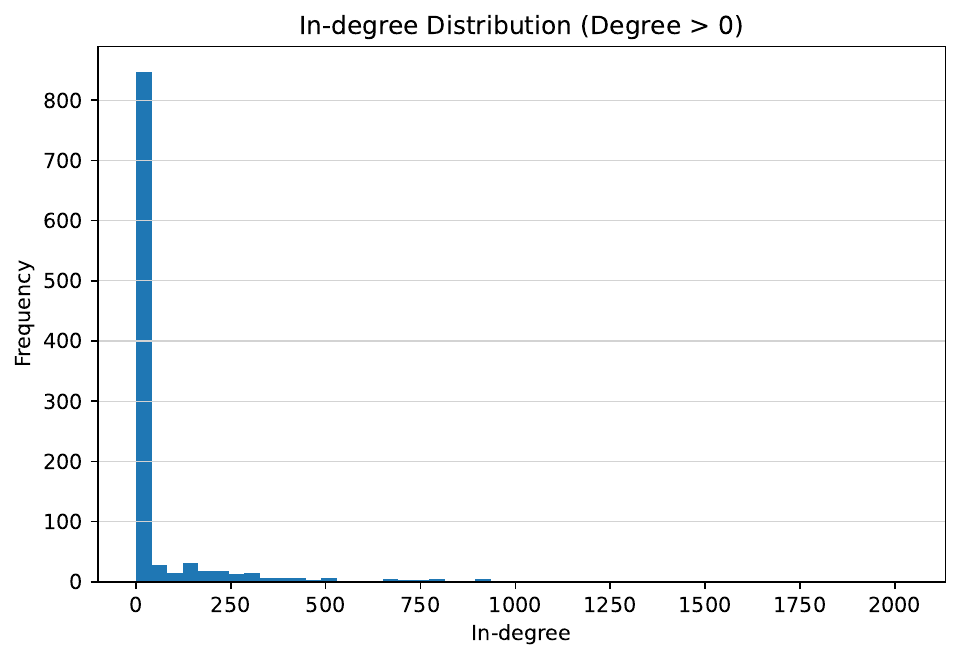}
    \end{subfigure}
    \hfill
    \begin{subfigure}[t]{0.49\textwidth}
        \centering
        \includegraphics[width=\textwidth]{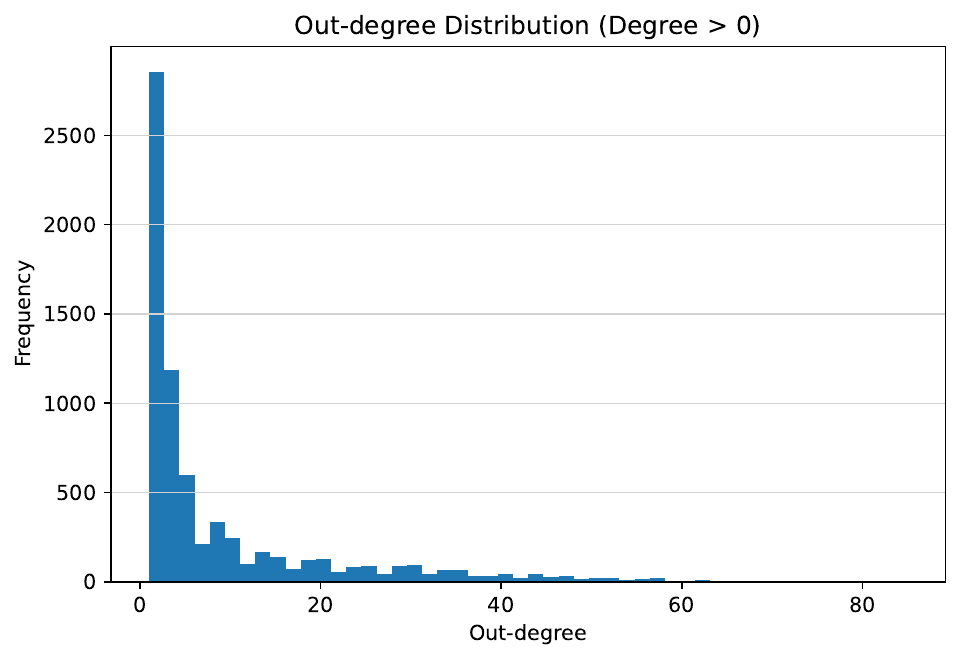}
    \end{subfigure}
    \caption{Histograms of in-degrees (left) and out-degrees (right) for the social media comment network.}
    \label{fig:degree}
\end{figure}

We further examine the network structure by plotting the histograms of in-degrees (number of followers per user) and out-degrees (number of followees per user) in Figure~\ref{fig:degree}. For clarity, nodes with zero degree are excluded from the plots, as these are primarily concentrated in the in-degree distribution. This exclusion highlights that most users actively follow others but are not themselves followed. The highly skewed in-degree distribution in Figure~\ref{fig:degree} reveals a small subset of users with exceptionally large numbers of followers, underscoring the presence of opinion leaders within the network.

\subsection{Performance Evaluation and Comparison}

To assess predictive performance, we employ a repeated random subsampling approach: 80\% of observations are used for training and 20\% for testing, with this partitioning repeated 100 times. Predictive accuracy is measured by the area under the ROC curve (AUC) on the test sets, comparing our network logistic regression (NLR) model against three alternatives: regression with network cohesion (RNC; \citealt{li2019prediction}), Simple Graph Convolution (SGC), and standard logistic regression (LR). Consistent with established practices in graph classification \citep{kipf2017semi, wu2019simplifying}, we set the propagation order to $K=2$ for both NLR and SGC.

Table~\ref{tab:auc} presents the mean AUC, standard error, and 95\% confidence intervals across the 100 test sets. NLR achieves the highest mean AUC of 93.50\% (SE = 0.52), outperforming RNC (92.71\%), SGC (92.40\%), and LR (89.82\%). Importantly, the lower bound of NLR's confidence interval (93.40\%) exceeds the upper bounds of all competing methods, indicating statistically significant improvements at conventional significance levels. The small standard error further highlights NLR's robustness and stability across resampling iterations.

\subsection{Methodological Comparisons and Insights}

The superior performance of network-based methods over standard logistic regression
%(mean AUC: NLR 93.50\%, RNC 92.71\%, SGC 92.40\% vs.\ LR 89.82\%)
underscores the importance of incorporating network structure into predictive modeling. This consistent advantage demonstrates that the adjacency matrix $A$ effectively captures node interdependencies, thereby enhancing classification accuracy.

Our proposed NLR model achieves a modest yet consistent improvement over SGC (93.50\% vs.\ 92.40\%), despite similar computational complexity. This improvement is attributable to NLR's flexible parameterization: the learnable coefficients $\bm{\lambda}_k$ allow adaptive weighting of neighborhood influences across different orders, accommodating local heterogeneity in network effects that SGC's fixed propagation scheme cannot capture. We do not include GCN in direct comparisons, as \citet{wu2019simplifying} have shown that SGC attains comparable accuracy with substantially lower computational cost.

RNC's strong performance (92.71\%) highlights the value of node-specific intercepts for modeling individual heterogeneity. However, this flexibility incurs significant computational overhead, as RNC requires estimating $N$ additional parameters. In contrast, NLR achieves superior accuracy with greater parsimony, estimating only $\tilde{d}$ parameters.

Notably, NLR's propagation structure can be interpreted as generating enhanced feature representations for classification. This perspective suggests further performance gains may be possible through adaptive weighting schemes, such as the AdaBoost-type strategy proposed by \citet{zhou2025model}, which could dynamically optimize the contributions of different propagation orders.

\begin{table}[htbp]
\centering
\caption{Predictive performance comparison on social network data. Reported are mean AUC (\%), standard error (SE), and 95\% confidence intervals (CI) across 100 test sets.}
\label{tab:auc}
\begin{tabular}{lccc}
\hline
\textbf{Method} & \textbf{AUC (\%)} & \textbf{SE} & \textbf{95\% CI (\%)} \\
\hline
LR   & 89.82 & 0.65 & (89.69, 89.95) \\
RNC  & 92.71 & 0.67 & (92.58, 92.84) \\
SGC  & 92.40 & 0.56 & (92.29, 92.51) \\
NLR  & \textbf{93.50} & \textbf{0.52} & \textbf{(93.40, 93.60)} \\
\hline
\end{tabular}
\end{table}

\section{Conclusion} \label{sec:conclusion}

We introduce a unified network propagation regression framework that efficiently models higher-order peer influence in network-dependent data. Our approach offers theoretical guarantees for estimator consistency and asymptotic normality, and provides interpretable coefficients for network effects.
Compared to existing methods, our framework is robust, parsimonious, and computationally efficient, outperforming alternatives in both simulations and real-world social media sentiment analysis. These strengths make it broadly applicable across social, biological, and information sciences.

Future directions include adaptive propagation schemes, dynamic network extensions, and integration with deep learning for enhanced representation and interpretability.

\bibliography{reference}

\end{document}